\documentclass[aip,rsi,reprint,graphicx]{revtex4-2} 

\usepackage{xcolor}
\usepackage{graphicx}
\usepackage{dcolumn}
\usepackage{bm}
\usepackage{amsmath}

\begin{document}

\preprint{AIP/123-QED}

\title{Dynamical measurement of saturation vapor pressures below and above room temperature}

\author{Mohsen Salimi}
\affiliation{Department of Physics and Astronomy, Aarhus University, DK-8000 Aarhus C, Denmark}

\author{Andreas B. Pedersen}
\affiliation{Department of Physics and Astronomy, Aarhus University, DK-8000 Aarhus C, Denmark}

\author{John E. V. Andersen}
\affiliation{Department of Physics and Astronomy, Aarhus University, DK-8000 Aarhus C, Denmark}


\author{Henrik B. Pedersen}
\affiliation{Department of Physics and Astronomy, Aarhus University, DK-8000 Aarhus C, Denmark}

\author{Aur\'{e}lien Dantan}
\thanks{To whom correspondence should be addressed}
\email[]{dantan@phys.au.dk}
\affiliation{Department of Physics and Astronomy, Aarhus University, DK-8000 Aarhus C, Denmark}

\date{\today}

\begin{abstract}
We report on the implementation of a dynamical method for the determination---in an extended temperature range around room temperature---of the saturation vapor pressure and enthalpy of vaporization of low-volatility liquid substances. The method relies on isolating a \textit{precooled} substance in a heated chamber under static vacuum conditions and monitoring the chamber pressure as the sample slowly thermalizes to the chamber temperature. We apply the method to four reference substances---diethyl phthalate, 1-decanol, 1-heptanol, and 1-hexanol---and provide accurate data for their saturation vapor pressure and enthalpy of vaporization in the range from $-10$ to $35^\circ$C.
\end{abstract}



\maketitle

\section{Introduction}
\label{sec:introduction}

The saturation vapor pressure (SVP)---the pressure exerted by a vapor at thermodynamic equilibrium with its condensed phase in a closed system---characterizes the propensity of a liquid to evaporate and condense, and, as such, is of utmost importance for a wide range of fields within atmospheric, health, chemical and physical sciences~\cite{calvert1990,mackay2014,bilde2015,krishnasamy2021}. The saturation vapor pressure increases nonlinearly with temperature, as described e.g. by the Clausius-Clapeyron relation for an ideal gas~\cite{wark1988}. This nonlinear dependence, together with the requirements on substance purity and absolute pressure measurement accuracy, contributes to the practical challenges of determining the saturation vapor pressure of low-volatility substances~\cite{bilde2015}. Most of the methods to determine saturation vapor pressures, using e.g. ionization gauges or Knudsen evaporation cells, are indirect---requiring a careful calibration of a number of experimental parameters---as well as time-consuming, especially when low-valued vapor pressures are to be obtained over a broad temperature range.

We recently commissioned an instrument for the absolute determination of saturation vapor pressures (ASVAP) using a direct, definition-based approach, in which a substance at room temperature is isolated in a heated chamber under static vacuum conditions and the chamber pressure is monitored as the sample slowly thermalized to the chamber temperature~\cite{nielsen2024}. This dynamical method was successfully applied to a number of reference substances and fatty acid methyl esters~\cite{nielsen2024,salimi2025}, and allowed for fast ($\sim$hour) and accurate measurements of their saturation vapor pressure from around room temperature to 35$^\circ$C and in a pressure range of $10^{-2}$ to $10^3$ Pa. The vapor pressure measurements in the temperature range available also enabled the determination of their enthalpy (heat) of vaporization at or slightly above room temperature.

In this work we report on the implementation of this dynamical method in an extended temperature and pressure range. The temperature range is extended well below room temperature by adding a sample precooling stage to the ASVAP instrument and the pressure range is extended by adding a spinning rotor gauge pressure sensor. We apply the method to four reference substances---diethyl phthalate, 1-decanol, 1-heptanol and 1-hexanol---and provide accurate data for their saturation vapor pressure and enthalpy of vaporization in the range $-10$ to $35^\circ$C and in the pressure range $10^{-3}-10^2$ Pa.

The paper is organized as follows: Section~\ref{sec:setup} presents the upgraded experimental apparatus. In Sec.~\ref{sec:methods} the substances investigated and the dynamical method used for determining their saturation vapor pressure and enthalpy of vaporization are introduced. The measurement results are reported in Section~\ref{sec:results} together with the final thermodynamical data. Section~\ref{sec:conclusion} concludes.

\section{Experimental system}
\label{sec:setup}

\subsection{Experimental setup and measurement procedure}

\begin{figure}[h]
\includegraphics[width=\columnwidth]{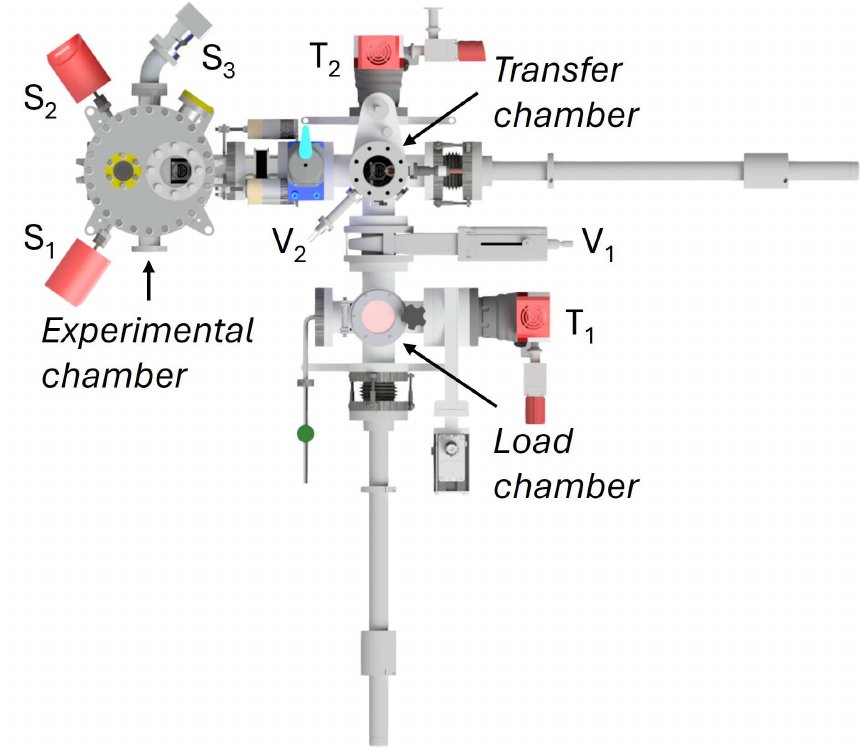}
\caption{Topview schematic of the ASVAP instrument. S$_1$, S$_2$, S$_3$: pressure sensors. T$_1$, T$_2$: valves. More details can be found in Ref.~\cite{nielsen2024}.}
\label{fig:topview}
\end{figure}

The ASVAP experimental setup has been described in detail in Ref.~\cite{nielsen2024} and a topview schematic of the main components is shown in Fig.~\ref{fig:topview}. In brief, the sample is inserted in the load chamber at room temperature and purified in low- or high-vacuum conditions for a certain period of time, depending on the vapor pressure. The sample is then transferred using the first transport arm to the room-temperature transfer chamber in which a high-vacuum (typically $\sim 10^{-5}-10^{-6}$ Pa) is achieved. In the transfer chamber the sample can be precooled using the precooling system described in the next section. 

When the desired sample temperature is reached, the sample can be transfered using the second transport arm to the experimental chamber, which is temperature stabilized at 35$^{\circ}$C for all the experiments reported here, and in which a high vacuum ($<10^{-6}$ Pa) has been realized prior to the insertion of the sample. A typical outgassing rate in the experimental chamber at 35$^\circ$C under static vacuum conditions is $\sim 10^{-3}$ Pa/h. Great care has also been paid to the thermal isolation of the experimental chamber in order to ensure high temperature stability and homogeneity. The experimental chamber temperature is controlled by a close-circuit chiller circulating liquid through the top and bottom flanges of the chamber, as well as in copper tubes along the chamber body. Finally, to avoid the presence of cold points, additional heating is applied to some of the flanges (e.g. valve S$_2$) and the ambient air around the chamber is actively heated to be higher by 0.1-0.3$^\circ$C than the set temperature of the chiller cooling liquid.

The measurement starts when valve V$_2$ is closed and the sample temperature starts rising as the sample holder is in contact with a large copper platform at the chamber temperature. As a result of evaporation and condensation of the sample isolated in the heated chamber the pressure in the chamber rises. The chamber pressure can be determined in an absolute manner in the range 5 mPa-1000 Pa by two capactive diaphragm sensors S$_1$ (BCEL7045 0.1 mbar, Edwards) for the range $5\times 10^{-3}-10$ Pa) and S$_2$ (CDG045D 10 mbar, INFICON) for the range $5\times 10^{-1}-1000$ Pa. Both sensors are temperature stabilized at 45$^\circ$ and have specified accuracies of 0.15\% and resolutions of 0.003\% from the manufacturers. In addition, a novel sensor S$_3$, a spinning rotor gauge (CaliTorr, ph-instruments), can be used to extend the pressure measurement range down to $5\times 10^{-5}$ Pa provided that a suitable, species-dependent calibration against S$_1$ is performed at pressures above $5\times 10^{-3}$ Pa.

\subsection{Sample cooling system}

\begin{figure}[h]
\includegraphics[width=0.65\columnwidth]{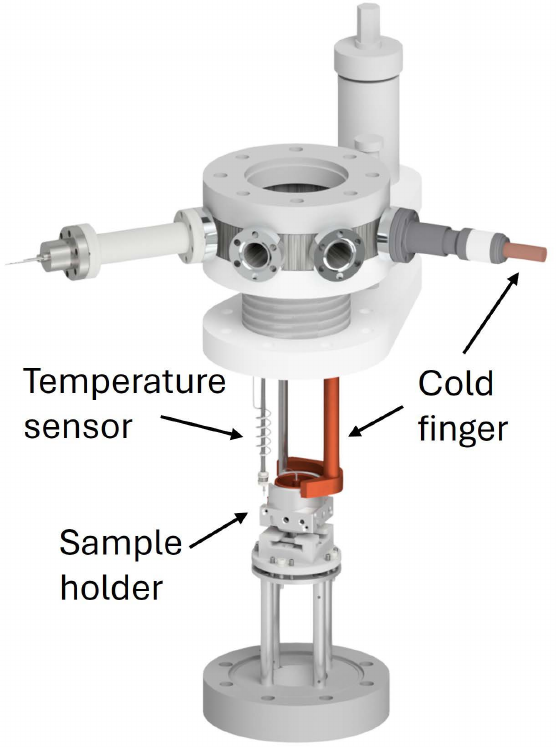}
\caption{Schematic of the novel sample cooling system in the transfer chamber.}
\label{fig:transfer_cooling}
\end{figure}

To extend the temperature range available for pressure measurements a system for precooling the sample in the transfer chamber has been installed. A schematic of the system is shown in Fig.~\ref{fig:transfer_cooling}. Once the sample holder has been transfered to its fixed platform in the transfer chamber, a cold copper finger can be brought down so that its lower half-ring shaped extremity is in good thermal contact with the base of the sample holder. The cold finger is precooled externally using liquid nitrogen. The temperature of the sample holder can be monitored using a movable temperature sensor that is also in contact with the sample holder base. Once the desired precooling temperature has been reached, the cold finger is lifted and the sample holder can be transfered to the experimental chamber.

\section{Methods}
\label{sec:methods}

 \subsection{Substance preparation}

In order to ascertain the applicability of the newly established dynamical method in an extended range around room temperature SVP measurements were performed for the same four reference substances used in Ref.~\cite{nielsen2024}. The selected substances are all liquid in the temperature ranges considered and are in order of increasing vapor 
pressure: diethyl phthalate (C$_{12}$H$_{14}$O$_4$), 1-decanol (C$_{10}$H$_{22}$O), 1-heptanol (C$_7$H$_{16}$O), and 
1-hexanol (C$_6$H$_{14}$O). Samples of these substances were purchased from Sigma-Aldrich/Merck and the certified purities, obtained with gas chromatography, were 
99.8\,\%, 99.6\,\%, 99.8\,\%, and 99.5\,\%, respectively.
  
Prior to the measurements, the samples were further purified
by repeated cycles of evacuation (active pumping and evaporative cooling) 
and reheating under static vacuum conditions (no vacuum pumping), 
whereby substances (e.g. water) with higher saturation vapor 
pressures initially dissolved in the sample gradually escape 
the sample. The initial amount of sample and pumping sequences/times were optimized for each substance with the aim of keeping enough sample for verifying the reproducibility of the measurements in the experimental chamber. For 1-hexanol, 5 mL of sample were pumped for 50 minutes with a backing pump in the load chamber and then pumped for 50 minutes with a turbo pump in the transfer chamber during precooling. For 1-heptanol, 4 mL of sample were pumped for 65 minutes with a backing pump in the load chamber, followed by 7 short (few seconds) pumping sequences with a turbo pump. After an additional 30 minutes pumping with the backing pump the sample was pumped for 1 hour with a turbo pump in the transfer chamber during precooling. For 1-decanol, 4 mL of sample were pumped for 1 hour with a backing pump in the load chamber, followed by 40 sequences of 1 minute pumping with a turbo pump and 1 hour pumping with a backing pump. The sample was then pumped for 1 hour with a turbo pump in the transfer chamber during precooling. For diethyl phtalate, 4 mL of sample were pumped for 1 hour with a backing pump and then 19 hours with a turbo pump in the load chamber, before being pumped for 1 hour with a turbo pump in the transfer chamber during precooling.

\subsection{Dynamical determination of the vapor pressure}

The dynamical measurement consists in monitoring the chamber pressure $p_V$ using one or several of the three available pressure sensors as a function of the sample temperature $T_L$, as the sample slowly thermalizes to the chamber temperature $T_V$. 
The determination of the SVP is performed using the statistical rate theory (SRT) for the particle flux during the evaporation and condensation processes~\cite{persad2010}, as exposed in Ref.~\cite{nielsen2024}. In brief, under steady state condition for the net particle flux, the chamber pressure can be written as
\begin{equation}
p_V=p_\textrm{sat}(T_L)\times f_\textrm{SRT}(T_L,T_V,\omega_i),
\label{eq:pV}
\end{equation}
where $p_\textrm{sat}(T_L)$ is the saturation vapor pressure of the substance at temperature $T_L$ and $f_\textrm{SRT}$ is a characteristic function that depends on the sample temperature $T_L$ and the chamber temperature $T_V$, as well as the vibrational mode frequencies $\omega_i$ of the molecules, where $i=1,..., \textrm{DOF}$, where DOF is the number of vibrational degrees of freedom of the molecules ($3N-6$, where $N$ is the number of atoms in the molecule).

\subsubsection{Saturation vapor pressure}

For an ideal gas the temperature dependence of the SVP is given by the Clausius-Clapeyron equation~\cite{wark1988}
\begin{equation}
\frac{dp_\textrm{sat}}{dT}=\frac{\Delta H_\textrm{vap}}{k_\textrm{B}N_\textrm{A}T^2}\times p_\textrm{sat},
\label{eq:dpsat}
\end{equation}
where $\Delta H_\textrm{vap}$ is the enthalpy of vaporization, $k_\textrm{B}$ is Boltzmann's constant and $N_\textrm{A}$ is Avogadro's number. Given the larger temperature range investigated in the present work as compared to Ref.~\cite{nielsen2024}, we are interested in taking into account the variation of the enthalpy of vaporization with temperature in a limited temperature range around a given (e.g. room) temperature $T^*$. These variations are related to the difference in the heat capacities of the gas and liquid phases, $C_p^{(g)}$ and $C_p^{(l)}$, by
\begin{equation}
\frac{d\Delta H_\textrm{vap}}{dT}=C_p^{(g)}-C_p^{(l)},
\end{equation}
which, as will be shown in the next section, can be well-approximated  in the range considered by a linear variation around $T^*$
\begin{equation}C_p^{(g)}-C_p^{(l)}\simeq \beta+\alpha(T-T^*),
\end{equation}
where $\beta$ and $\alpha$ are constants. Integrating this equation yields a close-to-linear variation of $\Delta H_\textrm{vap}$ around $T^*$ with a small quadratic correction
\begin{equation}
\Delta H_\textrm{vap}=\Delta H_\textrm{vap}^*+\beta(T-T^*)+\frac{1}{2}\alpha(T-T^*)^2,
\label{eq:DH}
\end{equation}
where $\Delta H_\textrm{vap}^*$ is the enthalpy of vaporization at $T^*$. Integrating Eq.~(\ref{eq:dpsat}) with the expression for $\Delta H_\textrm{vap}$  given by Eq.~(\ref{eq:DH}) yields an expected variation of the saturation vapor pressure with temperature given by
\begin{align}
\nonumber p_\textrm{sat}=&p^*_\textrm{sat}\exp\left[-\frac{\Delta H_\textrm{vap}^*}{k_\textrm{B}N_\textrm{A}}\left(\frac{1}{T}-\frac{1}{T^*}\right)\right.\\
\nonumber &-\frac{\beta}{k_\textrm{B}N_\textrm{A}}\left(1-\frac{T^*}{T}+\ln\frac{T^*}{T}\right)\\
&\left. +\frac{\alpha T^*}{2k_\textrm{B}N_\textrm{A}}\left(\frac{T}{T^*}-\frac{T^*}{T}+2\ln\frac{T^*}{T}\right)\right],
\label{eq:psatfinal}
\end{align}
where $p^*_\textrm{sat}$ is the saturation vapor pressure at temperature $T^*$.

\subsubsection{Heat capacities}

\begin{figure}[h]
\centering
\includegraphics[width=0.8\columnwidth]{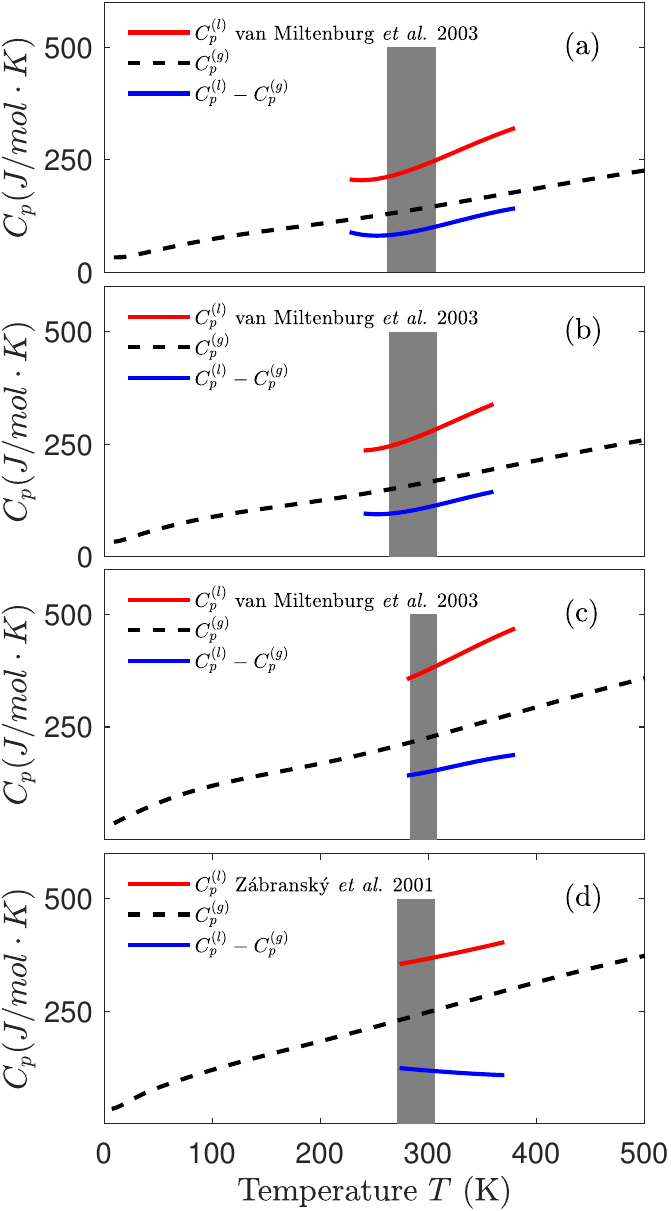}
\caption{Variation of the heat capacities of the liquid (red lines) and gaseous (black dashed lines) substances with temperature for (a) 1-hexanol, (b) 1-heptanol, (c) 1-decanol and (d) diethyl phthalate. $C_p^{(g)}$ was calculated using Eq.~(\ref{eq:Cpg}). The blue lines show their difference, $C_p^{(l)}-C_p^{(g)}$. The grey area indicates the temperature range relevant to this work.}
\label{fig:Cp}
\end{figure}

To estimate the approximate variation of the heat capacities in the temperature range considered, i.e. to determine $\alpha$ and $\beta$, we use the values reported in the literature for {\it liquid} 1-hexanol, 1-heptanol, 1-decanol~\cite{vanmiltenburg2003} and diethyl phthalate~\cite{zabransky2001}. The heat capacities at constant pressure for the {\it gaseous} substances are calculated from the vibrational frequencies of the molecules computed in Ref.~\cite{nielsen2024} as
\begin{equation}
C_p^{(g)}=4k_\textrm{B}N_\textrm{A}+k_\textrm{B}N_\textrm{A}D_e(T),
\label{eq:Cpg}
\end{equation}
where $D_e(T)$ is the effective DOF given by
\begin{equation}
D_e(T)=\sum_i \left(\frac{\hbar\omega_i}{k_\textrm{B}T}\right)^2\frac{e^{\hbar\omega_i/k_\textrm{B}T}}{\left(e^{\hbar\omega_i/k_\textrm{B}T}-1\right)^2}.
\label{eq:Dedef}
\end{equation}
The variation with temperature of the calculated heat capacities for the four substances considered are shown in Fig.~\ref{fig:Cp}. The values of $\alpha$ and $\beta$ resulting from fits of $C_p^{(g)}-C_p^{(l)}$ with a linear variation of the form $\beta+\alpha(T-T^*)$ in the temperature range considered around $T^*=25^\circ$ C are reported in Table~\ref{tab:Cp}. 

\begin{table*}[t!]
\caption{
Heat capacities $C_p^{(l)*}$ and $C_p^{(g)*}$ at $T^*=25^\circ$C for the four reference substances. Best fit values for $\beta$ and $\alpha$ resulting from a fit of $C_p^{(g)}-C_p^{(l)}$ with $\beta+\alpha(T-T^*)$ in the relevant temperature range. Physical and effective numbers of degrees of freedom (evaluated at $T^*$).
}
\begin{tabular}{l|rr|rr|rr}
\hline
Material              	& $C_p^{(l)*}$  (J/mol/K) & $C_p^{(g)*}$  (J/mol/K) & $\beta$  (J/mol/K) & $\alpha$ (J/mol/K$^2$) & $DOF$ & $D_e(T^*)$\\
\hline
1-hexanol         		& 240.7 & 143.9 & $-96.7$  & $-0.447$ & 57 & 13.3 \\
1-heptanol          	& 274.3 & 165.0 & $-109.2$ & $-0.445$  & 66 & 15.8\\
1-decanol 			& 375.1 & 225.4 & $-149.8$ & $-0.464$ & 93 & 23.1\\
Diethyl phthalate   	& 366.3 & 247.3 & $-119.0$  & 0.203 & 84 & 25.8\\
\hline
\end{tabular}
\label{tab:Cp}
\end{table*}

\subsubsection{Determination of $f_\textrm{SRT}$}

\begin{figure}[h]
\centering
\includegraphics[width=0.7\columnwidth]{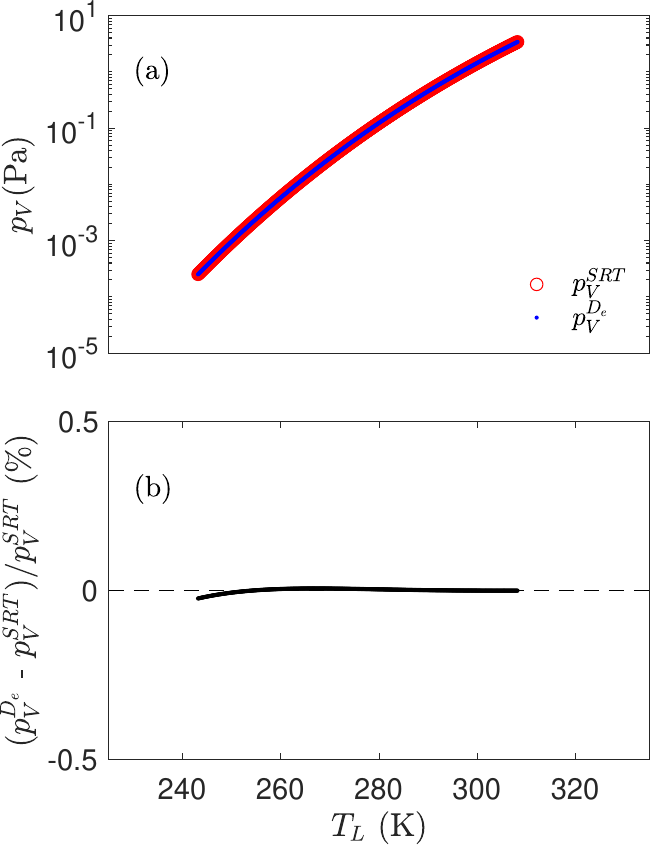}
\caption{(a) Variation with sample temperature $T_L$ of the chamber pressure $p_V$ predicted by Eq.~(\ref{eq:pV}) using the parametrization of $p_\textrm{sat}$ given by Eq.~(\ref{eq:psatfinal}) and the values $p^*_\textrm{sat}=1.2$ Pa and $\Delta H_\textrm{vap}^*=81.2$ kJ/mol reported in Ref.~\cite{pokorny2021}. The chamber temperature is $T_V=35^\circ$C. The blue dashed line indicate the result of a fit with Eq.~(\ref{eq:pVDe}), yielding an effective temperature $T_e=0.34T_L+0.66 T_V$ for the effective number of degrees of freedom. (b) Fit residuals.}
\label{fig:Deff}
\end{figure}

The function $f_\textrm{SRT}$ must be generally determined numerically from the steady state flux condition~\cite{nielsen2024} and relies on the explicit knowledge of the molecules' vibrational frequencies. However, for the substances and temperature and pressure ranges investigated here, it can be well-approximated by the analytical expression
\begin{equation}
p_V^{De}=p_\textrm{sat}(T_L)\times \exp\left[(D_e+4)\left(1-\frac{T_V}{T_L}\right)\right]\left(\frac{T_V}{T_L}\right)^{D_e+4},
\label{eq:pVDe}
\end{equation}
where $D_e$ represents the effective number of vibrational degrees of freedom ($0\leq D_e\leq\textrm{DOF}$), evaluated from Eq.~\ref{eq:Dedef} at an effective temperature $T_e=0.34T_L+0.66 T_V$, which can be determined from a fit of the predictions of the full SRT model with Eq.~(\ref{eq:pVDe}). This is illustrated in Fig.~\ref{fig:Deff} in the case of 1-decanol. For the substances investigated and in the range considered we checked that the results from fitting with the analytical expression Eq.~(\ref{eq:pVDe}) are indistinguishable from those obtained when fitting with the full SRT model of Ref.~\cite{nielsen2024}. The values of the effective number of degrees of freedom at $T^*$ are reported in Table~\ref{tab:Cp}. 

This strategy is advantageous in practice, since, once the vibrational frequencies of the substance investigated have been calculated, the closed form expression for the chamber pressure provided by Eq.~(\ref{eq:pVDe}) can be used for accurately fitting the data much more rapidly than using the complex SRT model expressions exposed in Ref.~\cite{nielsen2024}. When fitting the data with Eq.~(\ref{eq:pVDe}),  $T^*$ is chosen to be 25$^\circ$C, the values of $\alpha$ and $\beta$ are fixed to those calculated in the previous section (see Table~\ref{tab:Cp}), and $p^*_\textrm{sat}$ and $\Delta H_\textrm{vap}^*$ are free fitting parameters.

\subsubsection{Uncertainties}

The sample temperature is determined by a self-assembled sensor (type K, chromel-alumel), calibrated against a commercially available and accurate PT100 sensor (SE02, Pico Technology) with a final estimated accuracy $\delta T_L=0.1$ K. The specified pressure uncertainties for sensors S$_1$ and S$_2$ are $\delta p_\textrm{S$_1$}=3\times 10^{-4}+1.5\times 10^{-3}p_\textrm{S$_1$}$ and $\delta p_\textrm{S$_2$}=3\times 10^{-2}+1.5\times 10^{-3}p_\textrm{S$_2$}$. For the measurements with diethyl phthalate, the spinning rotor gauge sensor S$_3$ is calibrated against the absolute sensor S$_1$ in the range 0.2-0.3 Pa and its calibrated pressure readings in the whole measurement range (2 mPa-0.3 Pa) are subsequently used for the fitting with the effective SRT model (Eq.~(\ref{eq:pVDe})). To avoid potential systematic effects due the initial evaporation dynamics the first few minutes of the experimental data are not used for fitting with the steady state model predictions. The final uncertainties on $p^*_\textrm{sat}$ and $\Delta H_\textrm{vap}^*$ are calculated by fitting the data with Eq.~(\ref{eq:pVDe}) taking into account the previously mentioned temperature and pressure uncertainties, and are reported with a $2\sigma$ deviation.


\section{Experimental results}
\label{sec:results}

\subsection{Pressure measurement results}

\begin{figure}[h]
\includegraphics[width=\columnwidth]{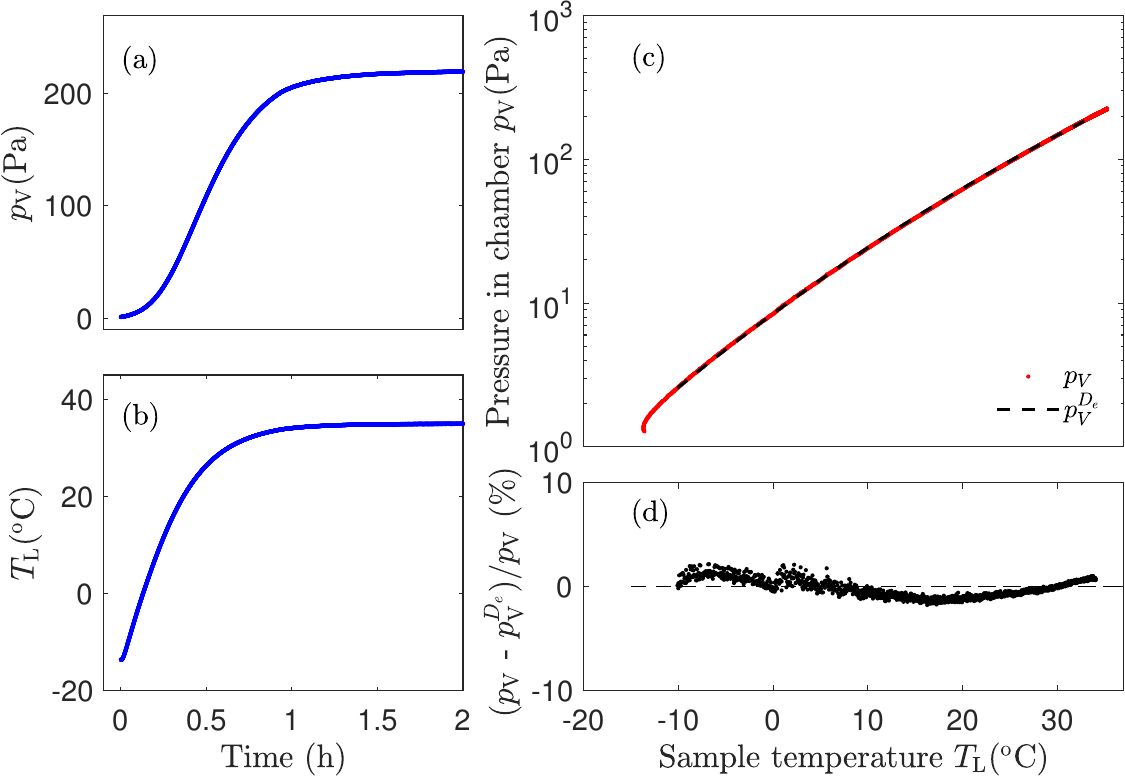}
\caption{Measurement results for 1-hexanol. (a,b) Variation of the chamber pressure $p_V$ and sample temperature $T_L$ as a function of time. (c) Chamber pressure $p_V$ as a function of sample temperature $T_L$. The black, dashed line shows the result of the fit with Eq.~(\ref{eq:pVDe}). (d) Fit residuals.}
\label{fig:hexanol}
\end{figure}

\begin{figure}[h]
\includegraphics[width=\columnwidth]{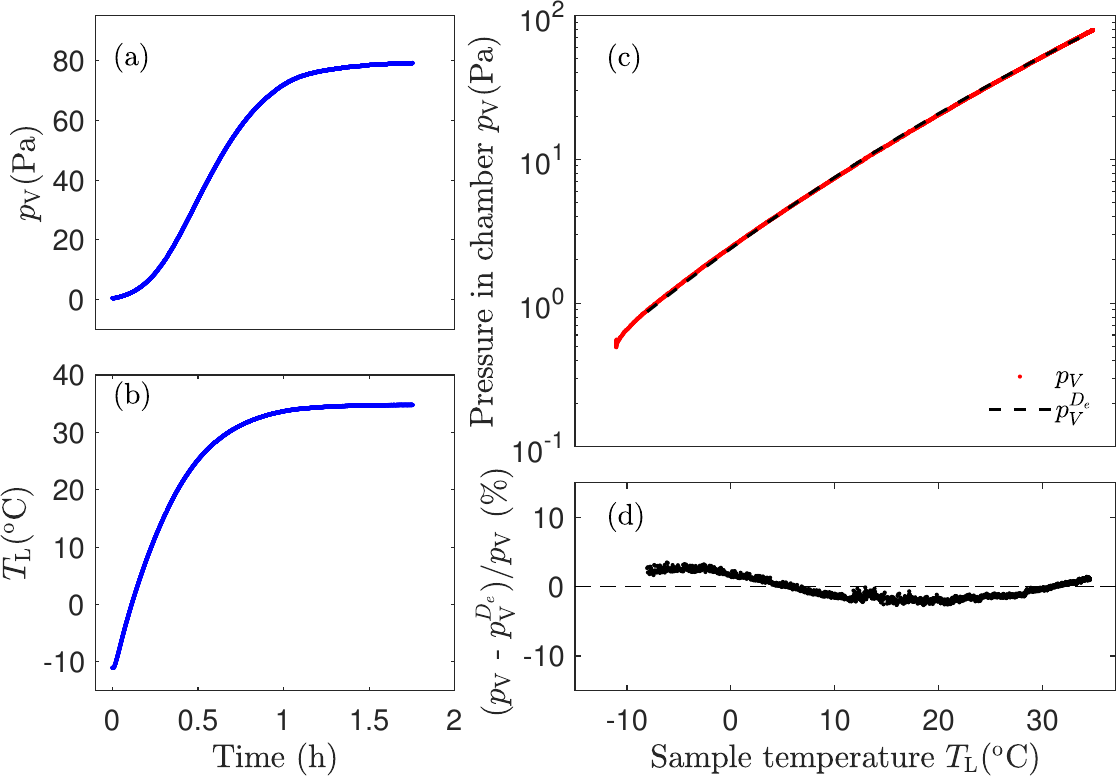}
\caption{Measurement results for 1-heptanol. (a,b) Variation of the chamber pressure $p_V$ and sample temperature $T_L$ as a function of time. (c) Chamber pressure $p_V$ as a function of sample temperature $T_L$. The black, dashed line shows the result of the fit with Eq.~(\ref{eq:pVDe}). (d) Fit residuals.}
\label{fig:heptanol}
\end{figure}

\begin{figure}[h]
\includegraphics[width=\columnwidth]{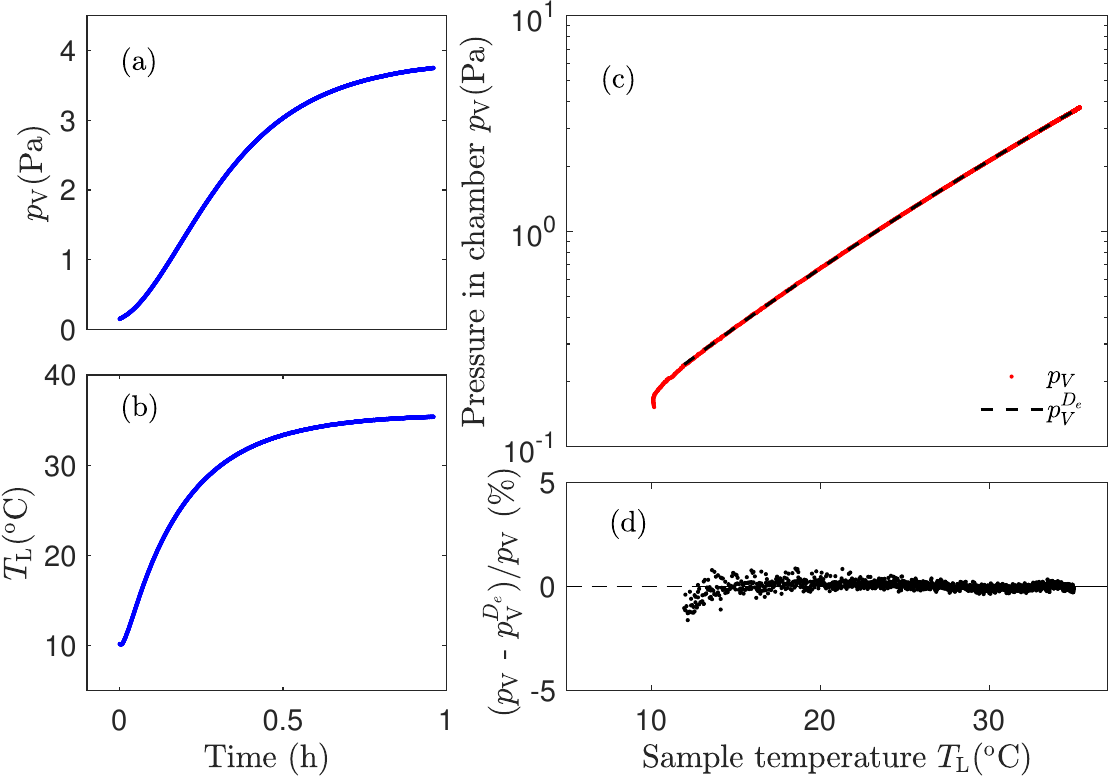}
\caption{Measurement results for 1-decanol. (a,b) Variation of the chamber pressure $p_V$ and sample temperature $T_L$ as a function of time. (c) Chamber pressure $p_V$ as a function of sample temperature $T_L$. The black, dashed line shows the result of the fit with Eq.~(\ref{eq:pVDe}). (d) Fit residuals.}
\label{fig:decanol}
\end{figure}

\begin{figure}[h]
\includegraphics[width=\columnwidth]{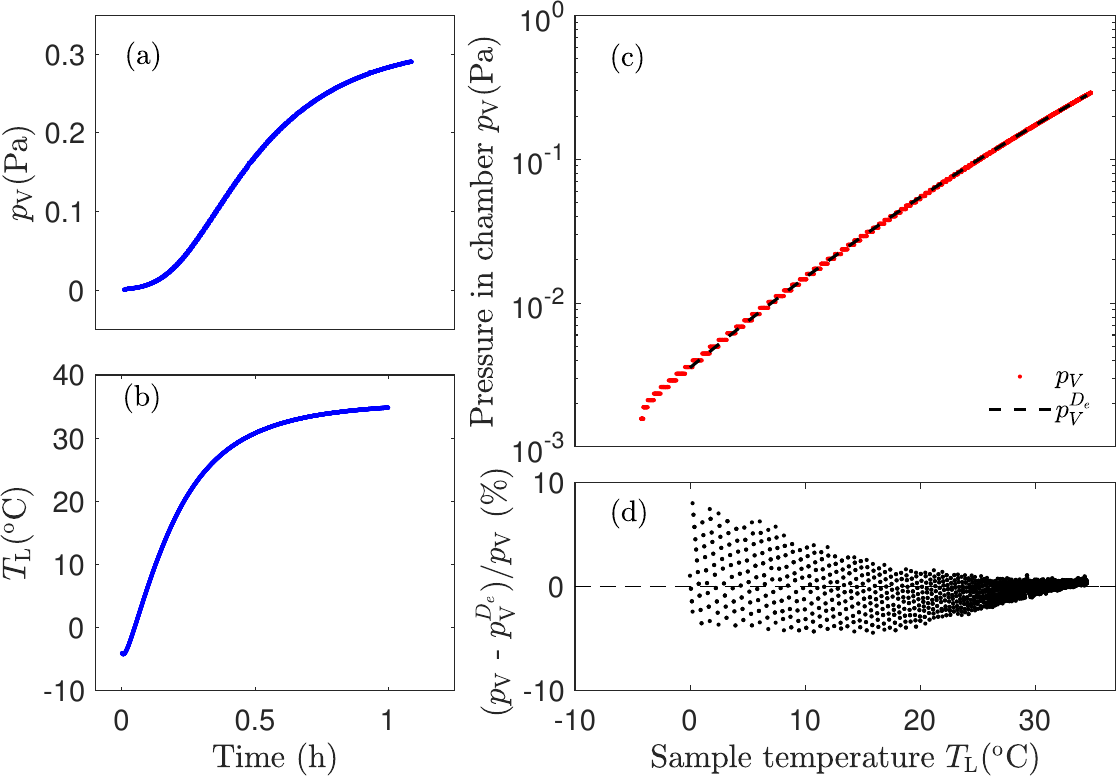}
\caption{Measurement results for diethyl phthalate. (a,b) Variation of the chamber pressure $p_V$ and sample temperature $T_L$ as a function of time. (c) Chamber pressure $p_V$ as a function of sample temperature $T_L$. The black, dashed line shows the result of the fit with Eq.~(\ref{eq:pVDe}). (d) Fit residuals.}
\label{fig:diethyl}
\end{figure}

The measured dynamical variations of the chamber pressure and the sample temperature are shown in Figs.~\ref{fig:hexanol}, \ref{fig:heptanol}, \ref{fig:decanol} and \ref{fig:diethyl}. Note, that in the case of 1-decanol, which has a melting temperature of $\sim 7^\circ$C, the liquid sample was only precooled down to 10$^\circ$C. An exponential rise in the temperature with a time constant of typically about 1000 seconds is typically observed for all substances (subfigures (b)). The pressure in the chamber is also observed to follow adiabatically the sample temperature after the first few minutes it takes for the evaporation and condensation processes to reach steady state (subfigures (a)). Subfigures (c) show the extracted $p_V$ versus $T_L$ curve, which displays the expected close-to-exponential increase in pressure with temperature. The results of the fit with the effective SRT model (Eq.~\ref{eq:pVDe}) are displayed as the black dashed lines in the temperature range chosen for fitting in subfigures (c) and the fit residuals are shown in subfigures (d). The results of the fit with the effective SRT model in which the expected temperature dependence of the enthalpy of vaporization has been included are observed to reproduce well the measured data over the whole temperature range. Let us remark that the result of fits with $\beta$ and $\alpha$ left as additional free parameters typically give similarly good agreement with the model, but do not allow to recover the expected temperature dependence of the enthalpy of vaporization in the range considered. This is why, in order to avoid potential systematic errors due to overparametrization in the estimation of $p^*_\textrm{sat}$ and $\Delta H^*_\textrm{vap}$, we fix the values of $\alpha$ and $\beta$ to those given in Table~\ref{tab:Cp}.

\subsection{Thermodynamical data}

\begin{figure}
\includegraphics[width=\columnwidth]{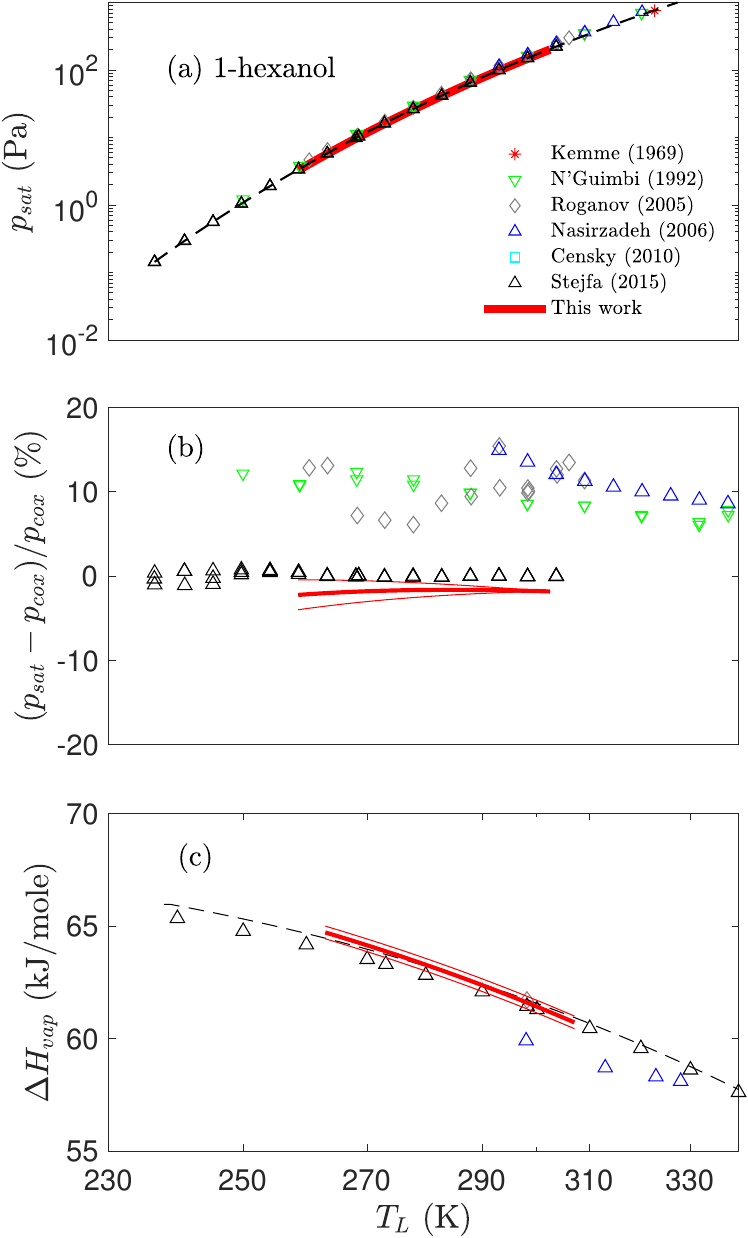}
\caption{Thermodynamical data for 1-hexanol. (a) Determined saturation vapor pressure $p_\textrm{sat}$ as a function of temperature (thick red line). (b) Relative deviation of $p_\textrm{sat}$ determined in this work with respect to the Cox parametrization of Ref.~\cite{stejfa2015}. (c) Determined enthalpy of vaporization $\Delta H_\textrm{vap}$ as a function of temperature (thick red line). The thin red lines show the $\pm2\sigma$ confidence interval. In (a) and (c) the black dashed line shows the Cox parametrization of Ref.~\cite{stejfa2015}. The symbols show the results of Refs.~\cite{kemme1969,nguimbi1992,roganov2005,nasirzadeh2006,censky2010,stejfa2015}.}
\label{fig:hexanola}
\end{figure}

\begin{figure}
\includegraphics[width=\columnwidth]{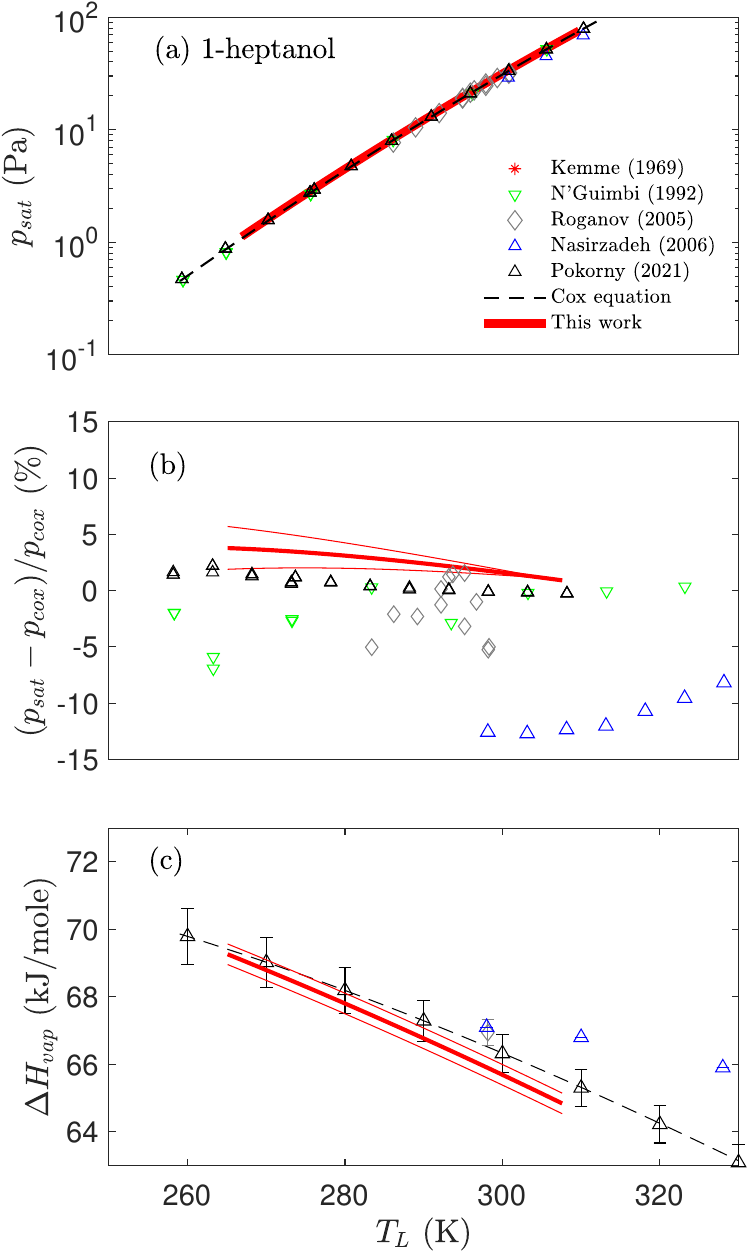}
\caption{Thermodynamical data for 1-heptanol. (a) Determined saturation vapor pressure $p_\textrm{sat}$ as a function of temperature (thick red line). (b) Relative deviation of $p_\textrm{sat}$ determined in this work with respect to the Cox parametrization of Ref.~\cite{pokorny2021}. (c) Determined enthalpy of vaporization $\Delta H_\textrm{vap}$ as a function of temperature (thick red line). The thin red lines show the $\pm2\sigma$ confidence interval. In (a) and (c) the black dashed line shows the Cox parametrization of Ref.~\cite{pokorny2021}. The symbols show the results of Refs.~\cite{kemme1969,nguimbi1992,roganov2005,nasirzadeh2006,pokorny2021}.}
\label{fig:heptanola}
\end{figure}

\begin{figure}
\includegraphics[width=\columnwidth]{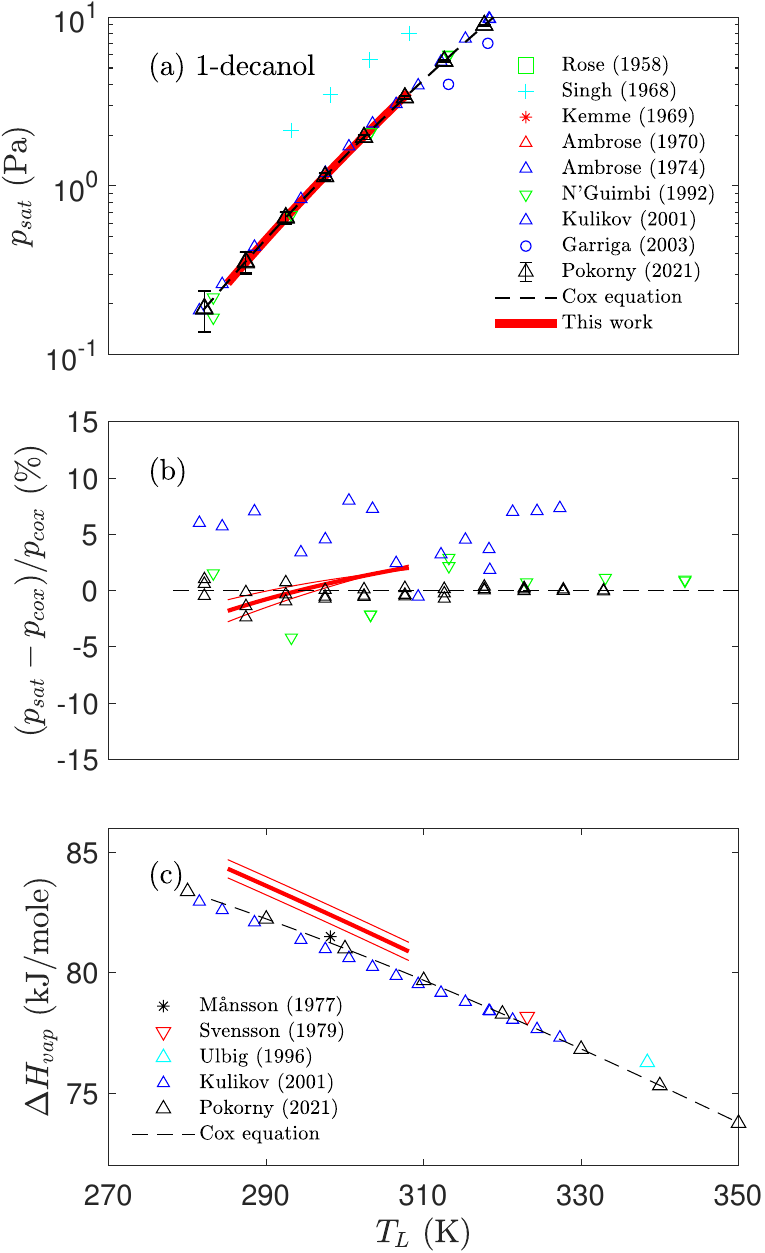}
\caption{Thermodynamical data for 1-decanol. (a) Determined saturation vapor pressure $p_\textrm{sat}$ as a function of temperature (thick red line). (b) Relative deviation of $p_\textrm{sat}$ determined in this work with respect to the Cox parametrization of Ref.~\cite{pokorny2021}. (c) Determined enthalpy of vaporization $\Delta H_\textrm{vap}$ as a function of temperature (thick red line). The thin red lines show the $\pm2\sigma$ confidence interval. In (a) and (c) the black dashed line shows the Cox parametrization of Ref.~\cite{pokorny2021}. The symbols show the results of Refs.~\cite{rose1958,singh1968,kemme1969,ambrose1970,ambrose1974,nguimbi1992,kulikov2001,garriga2003,pokorny2021,maansson1977,svensson1979,ulbig1996}.}
\label{fig:decanola}
\end{figure}

\begin{figure}
\includegraphics[width=\columnwidth]{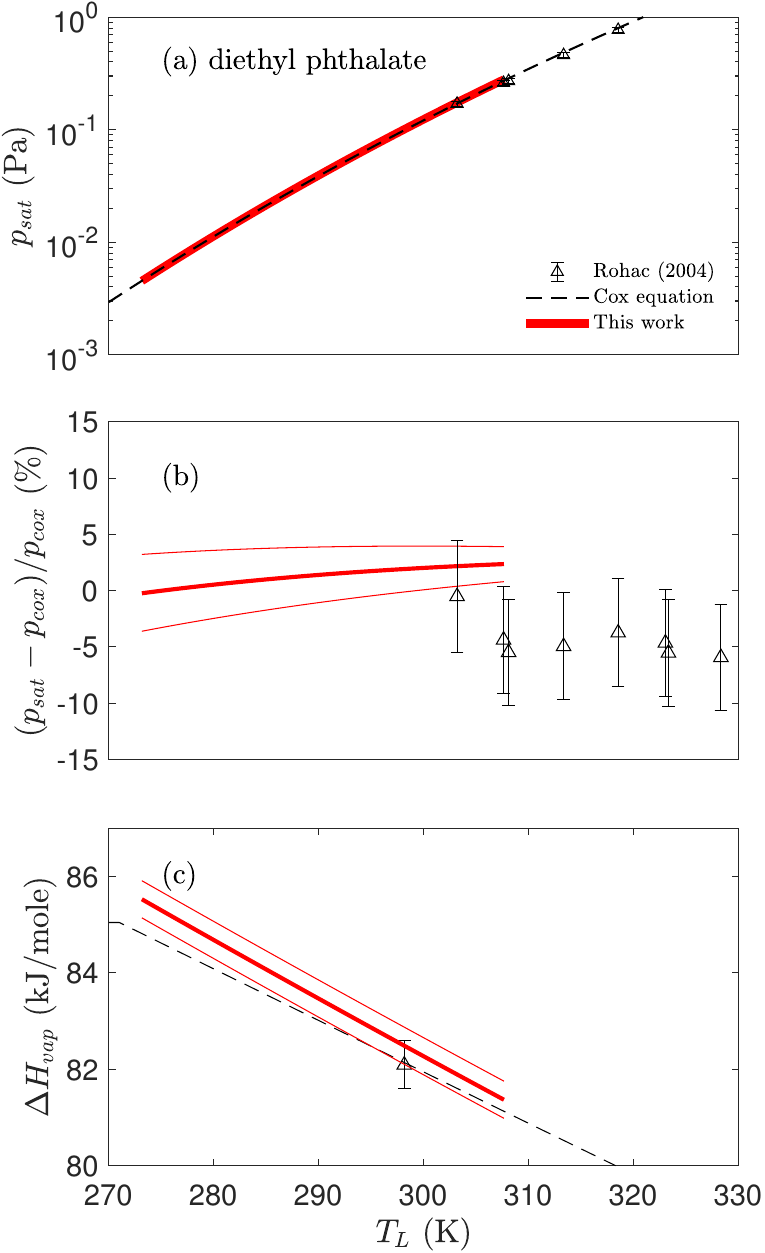}
\caption{Thermodynamical data for diethyl phthalate. (a) Determined saturation vapor pressure $p_\textrm{sat}$ as a function of temperature (thick red line). (b) Relative deviation of $p_\textrm{sat}$ determined in this work with respect to the Cox parametrization of Ref.~\cite{rohac2004}. (c) Determined enthalpy of vaporization $\Delta H_\textrm{vap}$ as a function of temperature (thick red line). The thin red lines show the $\pm2\sigma$ confidence interval. In (a) and (c) the black dashed line shows the Cox parametrization of Ref.~\cite{rohac2004}. The symbols show the results of Ref.~\cite{rohac2004}.}
\label{fig:diethyla}
\end{figure}

The derived values of $p_\textrm{sat}^*$ and $\Delta H_\textrm{vap}^*$ for each substance provide via Eq.~(\ref{eq:psatfinal}) the values of $p_\textrm{sat}$ and $\Delta H_\textrm{vap}$ in the temperature range investigated. These variations are reported as the thick red lines in Figs.~\ref{fig:hexanola}, \ref{fig:heptanola}, \ref{fig:decanola} and \ref{fig:diethyla}, the thin red lines indicating the $\pm 2\sigma$ standard deviation values. These values are also represented together with other values reported in the literature.

The saturation vapor pressures determined for 1-hexanol are in good agreement (slightly lower by 1-2\%) than those reported by Štejfa et al.~\cite{stejfa2015}, but significantly lower (by $\sim10\%$) than those reported by~\cite{nguimbi1992,roganov2005,nasirzadeh2006} in the temperature range considered. The values for the enthalpy of vaporization of 1-hexanol are consistent with those reported by Štejfa et al.~\cite{stejfa2015}. Similar observations can be made for 1-heptanol, for which the saturation vapor pressure values determined in this work are a few percent higher than those reported by Pokorn\'{y} et al.~\cite{pokorny2021}, N'Guimbi et al.~\cite{nguimbi1992} and Roganov et al.~\cite{roganov2005} and 10\% higher than those reported by Nasirzadeh et al.~\cite{nasirzadeh2006}. The enthalpy of vaporization values are consistent with those measured by Pokorn\'{y} et al.~\cite{pokorny2021}, and slightly lower by a few percent than those reported by Nasirzadeh et al.~\cite{nasirzadeh2006}. For 1-decanol the values of the saturation vapor pressure are in agreement with those reported by Pokorn\'{y} et al.~\cite{pokorny2021}, N'Guimbi et al.~\cite{nguimbi1992}, and lower by 5\% than those reported by Kulikov et al.~\cite{kulikov2001}. The values of the enthalpy of vaporization are slightly higher by 2\% than those reported by M{\aa}nsson et al.~\cite{maansson1977}, Svensson et al.~\cite{svensson1979}, Kulikov et al.~\cite{kulikov2001} and Pokorn\'{y} et al.~\cite{pokorny2021}. For diethyl phthalate the only existing measurements are to the best of our knowledge those of Roháč et al.~\cite{rohac2004} performed at temperatures above room temperature. The values of both the saturation vapor pressure and the enthalpy of vaporization are consistent with those at room temperature or above, and substantially extend the available data for this substance below room temperature. 

For completeness, the experimentally determined values for the saturation vapor pressure and the enthalpy of vaporization at room temperature are reported in Table~\ref{tab:p_sat_table}.

\begin{table*}[t!]
\caption{Experimentally determined vapor pressures at $T^*=25^\circ$C for the four substances. The specified uncertainties represent 95\,\% confidence intervals.}
\begin{tabular}{l|r|rr}
\hline
Material              	& Range ($\mathrm{^o}$C) &$p^{*}_\textrm{sat}$ (Pa)  &$\Delta H_{vap}^*$ (kJ/mole)    \\
\hline
1-hexanol          	& $-10$--34 & $98.4\pm 0.1$  & $61.6\pm 0.1$ \\
1-heptanol          	& $-8$--34.5 & $33.8\pm 0.1$  & $65.8\pm 0.2$ \\
1-decanol 		& 12-35 & $1.24\pm 0.01$  & $82.4\pm 0.2$ \\
Diethyl phthalate   	& 0-34.5 & $0.101\pm 0.002$  &  $82.5\pm 0.2$       \\
\hline
\end{tabular}
\label{tab:p_sat_table}
\end{table*}

\section{Conclusion}
\label{sec:conclusion}

We reported on the implementation of a fast and accurate dynamical method to determine the saturation vapor pressure and enthalpy of vaporization of low-volatility liquids by precooling the substance below room temperature, isolating it in a chamber heated to above room temperature and monitoring the chamber pressure as the substance adiabatically thermalizes to the chamber temperature. The addition of the sample precooling stage and of a more sensitive pressure sensor allowed for substantially extending both the temperature ($-10$ to $35^\circ$C) and pressure ($10^{-3}-10^2$ Pa) ranges of the apparatus described in Ref.~\cite{nielsen2024}. The method was successfully applied to four reference substances, providing thermodynamical data in good agreement with the literature for 1-hexanol, 1-heptanol and 1-decanol and new data for diethyl phthalate below room temperature.

Let us point out that the temperature range could straightforwardly be extended further in the low temperature range by improving the efficiency of the cooling transfer. The pressure range could also be extended either by fully exploiting the available range of the spinning rotor gauge (down to $5\times 10^{-5}$ Pa), barring a suitable calibration against an absolute pressure sensor can be performed at higher pressures, or by making use of the recently developed optomechanical absolute pressure sensors with demonstrated sensitivities at the $10^{-4}$ Pa level~\cite{salimi2024}.

Such extended pressure and temperature ranges are highly relevant for the determination of the saturation vapor pressure of substances relevant to aerosol particle formation in the atmosphere~\cite{bilde2015}. The ability to accurately measure low saturation vapor pressures below room temperature also opens for investigating the thermodynamics of solid substance evaporation and liquid-solid phase transitions, which is the focus of ongoing investigations.

\begin{acknowledgments}
We thank the mechanical workshop at Department of Physics and Astronomy, 
Aarhus University for support with the design and construction 
of the cooling system. We acknowledge financial support from the Villum foundation (grant no. 50229), the Novo Nordisk Foundation and Aahus Universitets Forskningsfond.
\end{acknowledgments}

\appendix*

\section*{Author declarations}
\subsection*{Conflict of interest}
The authors have no conflicts of interest.

\subsection*{Author contributions}

{\bf Mohsen Salimi:}
Data curation (lead);
Formal analysis (equal);
Investigation (lead);
Validation (equal);     
Visualization (equal); 
Writing - review and editing (equal);
{\bf Andreas B. Pedersen:}
Data curation (lead);
Formal analysis (equal);
Investigation (equal);
Validation (equal);     
Visualization (equal); 
Writing - review and editing (equal);
{\bf John E. V. Andersen:}
Investigation (equal);
Methodology (equal);
Validation (equal);     
Visualization (equal); 
Writing - review and editing (equal);
{\bf Henrik B. Pedersen:}
Conceptualization (lead);
Data curation (equal);
Formal analysis (lead);
Funding Acquisition (lead);
Investigation (lead);
Methodology (lead);
Project administration (lead);
Resources (lead);
Software (lead);
Supervision (lead);    
Validation (lead);     
Visualization (lead); 
Writing - original draft (equal); 
Writing - review and editing (equal);
{\bf Aur\'{e}lien Dantan:}
Conceptualization (lead);
Data curation (equal);
Formal analysis (lead);
Funding Acquisition (lead);
Investigation (lead);
Methodology (lead);
Project administration (lead);
Resources (lead);
Software (equal);
Supervision (lead);    
Validation (lead);     
Visualization (equal); 
Writing - original draft (lead); 
Writing - review and editing (lead).


\section*{Data Availability Statement}
The data that support the findings of this study are available
from the corresponding author upon reasonable request.


\bibliography{ASVAP_cool_bib}

\begin{thebibliography}{28}%
\makeatletter
\providecommand \@ifxundefined [1]{%
 \@ifx{#1\undefined}
}%
\providecommand \@ifnum [1]{%
 \ifnum #1\expandafter \@firstoftwo
 \else \expandafter \@secondoftwo
 \fi
}%
\providecommand \@ifx [1]{%
 \ifx #1\expandafter \@firstoftwo
 \else \expandafter \@secondoftwo
 \fi
}%
\providecommand \natexlab [1]{#1}%
\providecommand \enquote  [1]{``#1''}%
\providecommand \bibnamefont  [1]{#1}%
\providecommand \bibfnamefont [1]{#1}%
\providecommand \citenamefont [1]{#1}%
\providecommand \href@noop [0]{\@secondoftwo}%
\providecommand \href [0]{\begingroup \@sanitize@url \@href}%
\providecommand \@href[1]{\@@startlink{#1}\@@href}%
\providecommand \@@href[1]{\endgroup#1\@@endlink}%
\providecommand \@sanitize@url [0]{\catcode `\\12\catcode `\$12\catcode
  `\&12\catcode `\#12\catcode `\^12\catcode `\_12\catcode `\%12\relax}%
\providecommand \@@startlink[1]{}%
\providecommand \@@endlink[0]{}%
\providecommand \url  [0]{\begingroup\@sanitize@url \@url }%
\providecommand \@url [1]{\endgroup\@href {#1}{\urlprefix }}%
\providecommand \urlprefix  [0]{URL }%
\providecommand \Eprint [0]{\href }%
\providecommand \doibase [0]{https://doi.org/}%
\providecommand \selectlanguage [0]{\@gobble}%
\providecommand \bibinfo  [0]{\@secondoftwo}%
\providecommand \bibfield  [0]{\@secondoftwo}%
\providecommand \translation [1]{[#1]}%
\providecommand \BibitemOpen [0]{}%
\providecommand \bibitemStop [0]{}%
\providecommand \bibitemNoStop [0]{.\EOS\space}%
\providecommand \EOS [0]{\spacefactor3000\relax}%
\providecommand \BibitemShut  [1]{\csname bibitem#1\endcsname}%
\let\auto@bib@innerbib\@empty
\bibitem [{\citenamefont {Calvert}(1990)}]{calvert1990}%
  \BibitemOpen
  \bibfield  {author} {\bibinfo {author} {\bibfnamefont {J.~G.}\ \bibnamefont
  {Calvert}},\ }\bibfield  {title} {\enquote {\bibinfo {title} {Glossary of
  atmospheric chemistry terms (recommendations 1990)},}\ }\href
  {https://doi.org/doi:10.1351/pac199062112167} {\bibfield  {journal} {\bibinfo
   {journal} {Pure Appl. Chem.}\ }\textbf {\bibinfo {volume} {62}},\ \bibinfo
  {pages} {2167--2219} (\bibinfo {year} {1990})}\BibitemShut {NoStop}%
\bibitem [{\citenamefont {Mackay}\ and\ \citenamefont {van
  Wesenbeeck}(2014)}]{mackay2014}%
  \BibitemOpen
  \bibfield  {author} {\bibinfo {author} {\bibfnamefont {D.}~\bibnamefont
  {Mackay}}\ and\ \bibinfo {author} {\bibfnamefont {I.}~\bibnamefont {van
  Wesenbeeck}},\ }\bibfield  {title} {\enquote {\bibinfo {title} {Correlation
  of chemical evaporation rate with vapor pressure},}\ }\href
  {https://doi.org/10.1021/es5029074} {\bibfield  {journal} {\bibinfo
  {journal} {Environ. Sci. Technol.}\ }\textbf {\bibinfo {volume} {48}},\
  \bibinfo {pages} {10259--10263} (\bibinfo {year} {2014})},\ \bibinfo {note}
  {pMID: 25105222},\ \Eprint
  {https://arxiv.org/abs/https://doi.org/10.1021/es5029074}
  {https://doi.org/10.1021/es5029074} \BibitemShut {NoStop}%
\bibitem [{\citenamefont {Bilde}\ \emph {et~al.}(2015)\citenamefont {Bilde},
  \citenamefont {Barsanti}, \citenamefont {Booth}, \citenamefont {Cappa},
  \citenamefont {Donahue}, \citenamefont {Emanuelsson}, \citenamefont
  {McFiggans}, \citenamefont {Krieger}, \citenamefont {Marcolli}, \citenamefont
  {Topping}, \citenamefont {Ziemann}, \citenamefont {Barley}, \citenamefont
  {Clegg}, \citenamefont {Dennis-Smither}, \citenamefont {Hallquist},
  \citenamefont {Hallquist}, \citenamefont {Khlystov}, \citenamefont {Kulmala},
  \citenamefont {Mogensen}, \citenamefont {Percival}, \citenamefont {Pope},
  \citenamefont {Reid}, \citenamefont {Ribeiro~da Silva}, \citenamefont
  {Rosenoern}, \citenamefont {Salo}, \citenamefont {Soonsin}, \citenamefont
  {Yli-Juuti}, \citenamefont {Prisle}, \citenamefont {Pagels}, \citenamefont
  {Rarey}, \citenamefont {Zardini},\ and\ \citenamefont
  {Riipinen}}]{bilde2015}%
  \BibitemOpen
  \bibfield  {author} {\bibinfo {author} {\bibfnamefont {M.}~\bibnamefont
  {Bilde}}, \bibinfo {author} {\bibfnamefont {K.}~\bibnamefont {Barsanti}},
  \bibinfo {author} {\bibfnamefont {M.}~\bibnamefont {Booth}}, \bibinfo
  {author} {\bibfnamefont {C.~D.}\ \bibnamefont {Cappa}}, \bibinfo {author}
  {\bibfnamefont {N.~M.}\ \bibnamefont {Donahue}}, \bibinfo {author}
  {\bibfnamefont {E.~U.}\ \bibnamefont {Emanuelsson}}, \bibinfo {author}
  {\bibfnamefont {G.}~\bibnamefont {McFiggans}}, \bibinfo {author}
  {\bibfnamefont {U.~K.}\ \bibnamefont {Krieger}}, \bibinfo {author}
  {\bibfnamefont {C.}~\bibnamefont {Marcolli}}, \bibinfo {author}
  {\bibfnamefont {D.}~\bibnamefont {Topping}}, \bibinfo {author} {\bibfnamefont
  {P.}~\bibnamefont {Ziemann}}, \bibinfo {author} {\bibfnamefont
  {M.}~\bibnamefont {Barley}}, \bibinfo {author} {\bibfnamefont
  {S.}~\bibnamefont {Clegg}}, \bibinfo {author} {\bibfnamefont
  {B.}~\bibnamefont {Dennis-Smither}}, \bibinfo {author} {\bibfnamefont
  {M.}~\bibnamefont {Hallquist}}, \bibinfo {author} {\bibfnamefont {{\AA}.~M.}\
  \bibnamefont {Hallquist}}, \bibinfo {author} {\bibfnamefont {A.}~\bibnamefont
  {Khlystov}}, \bibinfo {author} {\bibfnamefont {M.}~\bibnamefont {Kulmala}},
  \bibinfo {author} {\bibfnamefont {D.}~\bibnamefont {Mogensen}}, \bibinfo
  {author} {\bibfnamefont {C.~J.}\ \bibnamefont {Percival}}, \bibinfo {author}
  {\bibfnamefont {F.}~\bibnamefont {Pope}}, \bibinfo {author} {\bibfnamefont
  {J.~P.}\ \bibnamefont {Reid}}, \bibinfo {author} {\bibfnamefont {M.~A.~V.}\
  \bibnamefont {Ribeiro~da Silva}}, \bibinfo {author} {\bibfnamefont
  {T.}~\bibnamefont {Rosenoern}}, \bibinfo {author} {\bibfnamefont
  {K.}~\bibnamefont {Salo}}, \bibinfo {author} {\bibfnamefont {V.~P.}\
  \bibnamefont {Soonsin}}, \bibinfo {author} {\bibfnamefont {T.}~\bibnamefont
  {Yli-Juuti}}, \bibinfo {author} {\bibfnamefont {N.~L.}\ \bibnamefont
  {Prisle}}, \bibinfo {author} {\bibfnamefont {J.}~\bibnamefont {Pagels}},
  \bibinfo {author} {\bibfnamefont {J.}~\bibnamefont {Rarey}}, \bibinfo
  {author} {\bibfnamefont {A.~A.}\ \bibnamefont {Zardini}},\ and\ \bibinfo
  {author} {\bibfnamefont {I.}~\bibnamefont {Riipinen}},\ }\bibfield  {title}
  {\enquote {\bibinfo {title} {Saturation vapor pressures and transition
  enthalpies of low-volatility organic molecules of atmospheric relevance: From
  dicarboxylic acids to complex mixtures},}\ }\href
  {https://doi.org/10.1021/cr5005502} {\bibfield  {journal} {\bibinfo
  {journal} {Chem. Rev.}\ }\textbf {\bibinfo {volume} {115}},\ \bibinfo {pages}
  {4115--4156} (\bibinfo {year} {2015})},\ \bibinfo {note} {pMID: 25929792},\
  \Eprint {https://arxiv.org/abs/https://doi.org/10.1021/cr5005502}
  {https://doi.org/10.1021/cr5005502} \BibitemShut {NoStop}%
\bibitem [{\citenamefont {Krishnasamy}\ and\ \citenamefont
  {Bukkarapu}(2021)}]{krishnasamy2021}%
  \BibitemOpen
  \bibfield  {author} {\bibinfo {author} {\bibfnamefont {A.}~\bibnamefont
  {Krishnasamy}}\ and\ \bibinfo {author} {\bibfnamefont {K.~R.}\ \bibnamefont
  {Bukkarapu}},\ }\bibfield  {title} {\enquote {\bibinfo {title} {A
  comprehensive review of biodiesel property prediction models for combustion
  modeling studies},}\ }\href
  {https://doi.org/https://doi.org/10.1016/j.fuel.2021.121085} {\bibfield
  {journal} {\bibinfo  {journal} {Fuel}\ }\textbf {\bibinfo {volume} {302}},\
  \bibinfo {pages} {121085} (\bibinfo {year} {2021})}\BibitemShut {NoStop}%
\bibitem [{\citenamefont {Wark}(1988)}]{wark1988}%
  \BibitemOpen
  \bibfield  {author} {\bibinfo {author} {\bibfnamefont {K.}~\bibnamefont
  {Wark}},\ }\href@noop {} {\emph {\bibinfo {title} {"Generalized Thermodynamic
  Relationships." Thermodynamics (5th ed.)}}}\ (\bibinfo  {publisher} {New
  York, NY: McGraw-Hill, Inc.},\ \bibinfo {year} {1988})\BibitemShut {NoStop}%
\bibitem [{\citenamefont {Nielsen}\ \emph {et~al.}(2024)\citenamefont
  {Nielsen}, \citenamefont {Salimi}, \citenamefont {Andersen}, \citenamefont
  {Elm}, \citenamefont {Dantan},\ and\ \citenamefont {Pedersen}}]{nielsen2024}%
  \BibitemOpen
  \bibfield  {author} {\bibinfo {author} {\bibfnamefont {R.~V.}\ \bibnamefont
  {Nielsen}}, \bibinfo {author} {\bibfnamefont {M.}~\bibnamefont {Salimi}},
  \bibinfo {author} {\bibfnamefont {J.~V.}\ \bibnamefont {Andersen}}, \bibinfo
  {author} {\bibfnamefont {J.}~\bibnamefont {Elm}}, \bibinfo {author}
  {\bibfnamefont {A.}~\bibnamefont {Dantan}},\ and\ \bibinfo {author}
  {\bibfnamefont {H.~B.}\ \bibnamefont {Pedersen}},\ }\bibfield  {title}
  {\enquote {\bibinfo {title} {A new setup for measurements of absolute
  saturation vapor pressures using a dynamical method: Experimental concept and
  validation},}\ }\href {https://doi.org/10.1063/5.0215176} {\bibfield
  {journal} {\bibinfo  {journal} {Rev. Sci. Instr.}\ }\textbf {\bibinfo
  {volume} {95}},\ \bibinfo {pages} {065007} (\bibinfo {year}
  {2024})}\BibitemShut {NoStop}%
\bibitem [{\citenamefont {Salimi}\ \emph {et~al.}(2025)\citenamefont {Salimi},
  \citenamefont {Nielsen}, \citenamefont {Elm}, \citenamefont {Dantan},\ and\
  \citenamefont {Pedersen}}]{salimi2025}%
  \BibitemOpen
  \bibfield  {author} {\bibinfo {author} {\bibfnamefont {M.}~\bibnamefont
  {Salimi}}, \bibinfo {author} {\bibfnamefont {R.~V.}\ \bibnamefont {Nielsen}},
  \bibinfo {author} {\bibfnamefont {J.}~\bibnamefont {Elm}}, \bibinfo {author}
  {\bibfnamefont {A.}~\bibnamefont {Dantan}},\ and\ \bibinfo {author}
  {\bibfnamefont {H.~B.}\ \bibnamefont {Pedersen}},\ }\bibfield  {title}
  {\enquote {\bibinfo {title} {Absolute saturation vapor pressures of three
  fatty acid methyl esters around room temperature},}\ }\href
  {https://doi.org/10.1021/acsomega.4c08095} {\bibfield  {journal} {\bibinfo
  {journal} {ACS Omega}\ }\textbf {\bibinfo {volume} {10}},\ \bibinfo {pages}
  {6671--6678} (\bibinfo {year} {2025})}\BibitemShut {NoStop}%
\bibitem [{\citenamefont {Persad}\ and\ \citenamefont
  {Ward}(2010)}]{persad2010}%
  \BibitemOpen
  \bibfield  {author} {\bibinfo {author} {\bibfnamefont {A.~H.}\ \bibnamefont
  {Persad}}\ and\ \bibinfo {author} {\bibfnamefont {C.~A.}\ \bibnamefont
  {Ward}},\ }\bibfield  {title} {\enquote {\bibinfo {title} {Statistical rate
  theory examination of ethanol evaporation},}\ }\href
  {https://doi.org/10.1021/jp100441m} {\bibfield  {journal} {\bibinfo
  {journal} {J. Phys. Chem. B}\ }\textbf {\bibinfo {volume} {114}},\ \bibinfo
  {pages} {6107--6116} (\bibinfo {year} {2010})},\ \bibinfo {note} {pMID:
  20405870},\ \Eprint {https://arxiv.org/abs/https://doi.org/10.1021/jp100441m}
  {https://doi.org/10.1021/jp100441m} \BibitemShut {NoStop}%
\bibitem [{\citenamefont {van Miltenburg}(2003)}]{vanmiltenburg2003}%
  \BibitemOpen
  \bibfield  {author} {\bibinfo {author} {\bibfnamefont {J.~C.}\ \bibnamefont
  {van Miltenburg}},\ }\bibfield  {title} {\enquote {\bibinfo {title} {Heat
  capacities and derived thermodynamic functions of 1-hexanol, 1-heptanol,
  1-octanol, and 1-decanol between 5 k and 390 k},}\ }\href
  {https://doi.org/10.1021/je0340856} {\bibfield  {journal} {\bibinfo
  {journal} {J. Chem. Eng. Data}\ }\textbf {\bibinfo {volume} {48}},\ \bibinfo
  {pages} {1323--1331} (\bibinfo {year} {2003})}\BibitemShut {NoStop}%
\bibitem [{\citenamefont {Zabransky}, \citenamefont {Ruzicka},\ and\
  \citenamefont {Domalski}(2001)}]{zabransky2001}%
  \BibitemOpen
  \bibfield  {author} {\bibinfo {author} {\bibfnamefont {M.}~\bibnamefont
  {Zabransky}}, \bibinfo {author} {\bibfnamefont {V.}~\bibnamefont {Ruzicka}},\
  and\ \bibinfo {author} {\bibfnamefont {E.~S.}\ \bibnamefont {Domalski}},\
  }\bibfield  {title} {\enquote {\bibinfo {title} {Heat capacity of liquids:
  Critical review and recommended values. supplement i},}\ }\href
  {https://doi.org/10.1063/1.1407866} {\bibfield  {journal} {\bibinfo
  {journal} {J. Phys. Chem. Ref. Data}\ }\textbf {\bibinfo {volume} {30}},\
  \bibinfo {pages} {1199--1689} (\bibinfo {year} {2001})}\BibitemShut {NoStop}%
\bibitem [{\citenamefont {Štejfa}\ \emph {et~al.}(2015)\citenamefont
  {Štejfa}, \citenamefont {Fulem}, \citenamefont {Růžička},\ and\
  \citenamefont {Matějka}}]{stejfa2015}%
  \BibitemOpen
  \bibfield  {author} {\bibinfo {author} {\bibfnamefont {V.}~\bibnamefont
  {Štejfa}}, \bibinfo {author} {\bibfnamefont {M.}~\bibnamefont {Fulem}},
  \bibinfo {author} {\bibfnamefont {K.}~\bibnamefont {Růžička}},\ and\
  \bibinfo {author} {\bibfnamefont {P.}~\bibnamefont {Matějka}},\ }\bibfield
  {title} {\enquote {\bibinfo {title} {Vapor pressures and thermophysical
  properties of selected hexenols and recommended vapor pressure for
  hexan-1-ol},}\ }\href
  {https://doi.org/https://doi.org/10.1016/j.fluid.2015.05.026} {\bibfield
  {journal} {\bibinfo  {journal} {Fluid Ph. Equilib.}\ }\textbf {\bibinfo
  {volume} {402}},\ \bibinfo {pages} {18--29} (\bibinfo {year}
  {2015})}\BibitemShut {NoStop}%
\bibitem [{\citenamefont {Pokorný}\ \emph {et~al.}(2021)\citenamefont
  {Pokorný}, \citenamefont {Štejfa}, \citenamefont {Klajmon}, \citenamefont
  {Fulem},\ and\ \citenamefont {Růžička}}]{pokorny2021}%
  \BibitemOpen
  \bibfield  {author} {\bibinfo {author} {\bibfnamefont {V.}~\bibnamefont
  {Pokorný}}, \bibinfo {author} {\bibfnamefont {V.}~\bibnamefont {Štejfa}},
  \bibinfo {author} {\bibfnamefont {M.}~\bibnamefont {Klajmon}}, \bibinfo
  {author} {\bibfnamefont {M.}~\bibnamefont {Fulem}},\ and\ \bibinfo {author}
  {\bibfnamefont {K.}~\bibnamefont {Růžička}},\ }\bibfield  {title}
  {\enquote {\bibinfo {title} {Vapor pressures and thermophysical properties of
  1-heptanol, 1-octanol, 1-nonanol, and 1-decanol: Data reconciliation and
  pc-saft modeling},}\ }\href {https://doi.org/10.1021/acs.jced.0c00878}
  {\bibfield  {journal} {\bibinfo  {journal} {J. Chem. Eng. Data}\ }\textbf
  {\bibinfo {volume} {66}},\ \bibinfo {pages} {805--821} (\bibinfo {year}
  {2021})},\ \Eprint
  {https://arxiv.org/abs/https://doi.org/10.1021/acs.jced.0c00878}
  {https://doi.org/10.1021/acs.jced.0c00878} \BibitemShut {NoStop}%
\bibitem [{\citenamefont {Roháč}\ \emph {et~al.}(2004)\citenamefont
  {Roháč}, \citenamefont {Růžička}, \citenamefont {Růžička},
  \citenamefont {Zaitsau}, \citenamefont {Kabo}, \citenamefont {Diky},\ and\
  \citenamefont {Aim}}]{rohac2004}%
  \BibitemOpen
  \bibfield  {author} {\bibinfo {author} {\bibfnamefont {V.}~\bibnamefont
  {Roháč}}, \bibinfo {author} {\bibfnamefont {K.}~\bibnamefont {Růžička}},
  \bibinfo {author} {\bibfnamefont {V.}~\bibnamefont {Růžička}}, \bibinfo
  {author} {\bibfnamefont {D.~H.}\ \bibnamefont {Zaitsau}}, \bibinfo {author}
  {\bibfnamefont {G.~J.}\ \bibnamefont {Kabo}}, \bibinfo {author}
  {\bibfnamefont {V.}~\bibnamefont {Diky}},\ and\ \bibinfo {author}
  {\bibfnamefont {K.}~\bibnamefont {Aim}},\ }\bibfield  {title} {\enquote
  {\bibinfo {title} {Vapour pressure of diethyl phthalate},}\ }\href
  {https://doi.org/https://doi.org/10.1016/j.jct.2004.07.025} {\bibfield
  {journal} {\bibinfo  {journal} {J. Chem. Thermodyn.}\ }\textbf {\bibinfo
  {volume} {36}},\ \bibinfo {pages} {929--937} (\bibinfo {year}
  {2004})}\BibitemShut {NoStop}%
\bibitem [{\citenamefont {N'Guimbi}\ \emph {et~al.}(1992)\citenamefont
  {N'Guimbi}, \citenamefont {Kasehgari}, \citenamefont {Mokbel},\ and\
  \citenamefont {Jose}}]{nguimbi1992}%
  \BibitemOpen
  \bibfield  {author} {\bibinfo {author} {\bibfnamefont {J.}~\bibnamefont
  {N'Guimbi}}, \bibinfo {author} {\bibfnamefont {H.}~\bibnamefont {Kasehgari}},
  \bibinfo {author} {\bibfnamefont {I.}~\bibnamefont {Mokbel}},\ and\ \bibinfo
  {author} {\bibfnamefont {J.}~\bibnamefont {Jose}},\ }\bibfield  {title}
  {\enquote {\bibinfo {title} {Tensions de vapeur d'alcools primaires dans le
  domaine 0,3 pa à 1,5 kpa},}\ }\href
  {https://doi.org/https://doi.org/10.1016/0040-6031(92)80100-B} {\bibfield
  {journal} {\bibinfo  {journal} {Thermochimica Acta}\ }\textbf {\bibinfo
  {volume} {196}},\ \bibinfo {pages} {367--377} (\bibinfo {year}
  {1992})}\BibitemShut {NoStop}%
\bibitem [{\citenamefont {Roganov}\ \emph {et~al.}(2005)\citenamefont
  {Roganov}, \citenamefont {Pisarev}, \citenamefont {Emel'yanenko},\ and\
  \citenamefont {Verevkin}}]{roganov2005}%
  \BibitemOpen
  \bibfield  {author} {\bibinfo {author} {\bibfnamefont {G.~N.}\ \bibnamefont
  {Roganov}}, \bibinfo {author} {\bibfnamefont {P.~N.}\ \bibnamefont
  {Pisarev}}, \bibinfo {author} {\bibfnamefont {V.~N.}\ \bibnamefont
  {Emel'yanenko}},\ and\ \bibinfo {author} {\bibfnamefont {S.~P.}\ \bibnamefont
  {Verevkin}},\ }\bibfield  {title} {\enquote {\bibinfo {title} {Measurement
  and prediction of thermochemical properties. improved benson-type increments
  for the estimation of enthalpies of vaporization and standard enthalpies of
  formation of aliphatic alcohols},}\ }\href
  {https://doi.org/10.1021/je049561m} {\bibfield  {journal} {\bibinfo
  {journal} {J. Chem. Eng. Data}\ }\textbf {\bibinfo {volume} {50}},\ \bibinfo
  {pages} {1114--1124} (\bibinfo {year} {2005})},\ \Eprint
  {https://arxiv.org/abs/https://doi.org/10.1021/je049561m}
  {https://doi.org/10.1021/je049561m} \BibitemShut {NoStop}%
\bibitem [{\citenamefont {Nasirzadeh}, \citenamefont {Neueder},\ and\
  \citenamefont {Kunz}(2006)}]{nasirzadeh2006}%
  \BibitemOpen
  \bibfield  {author} {\bibinfo {author} {\bibfnamefont {K.}~\bibnamefont
  {Nasirzadeh}}, \bibinfo {author} {\bibfnamefont {R.}~\bibnamefont
  {Neueder}},\ and\ \bibinfo {author} {\bibfnamefont {W.}~\bibnamefont
  {Kunz}},\ }\bibfield  {title} {\enquote {\bibinfo {title} {Vapor pressure
  determination of the aliphatic c5 to c8 1-alcohols},}\ }\href
  {https://doi.org/10.1021/je049600u} {\bibfield  {journal} {\bibinfo
  {journal} {J. Chem. Eng. Data}\ }\textbf {\bibinfo {volume} {51}},\ \bibinfo
  {pages} {7--10} (\bibinfo {year} {2006})},\ \Eprint
  {https://arxiv.org/abs/https://doi.org/10.1021/je049600u}
  {https://doi.org/10.1021/je049600u} \BibitemShut {NoStop}%
\bibitem [{\citenamefont {Kulikov}, \citenamefont {Verevkin},\ and\
  \citenamefont {Heintz}(2001)}]{kulikov2001}%
  \BibitemOpen
  \bibfield  {author} {\bibinfo {author} {\bibfnamefont {D.}~\bibnamefont
  {Kulikov}}, \bibinfo {author} {\bibfnamefont {S.~P.}\ \bibnamefont
  {Verevkin}},\ and\ \bibinfo {author} {\bibfnamefont {A.}~\bibnamefont
  {Heintz}},\ }\bibfield  {title} {\enquote {\bibinfo {title} {Enthalpies of
  vaporization of a series of aliphatic alcohols: Experimental results and
  values predicted by the eras-model},}\ }\href
  {https://doi.org/https://doi.org/10.1016/S0378-3812(01)00633-1} {\bibfield
  {journal} {\bibinfo  {journal} {Fluid Ph. Equilib.}\ }\textbf {\bibinfo
  {volume} {192}},\ \bibinfo {pages} {187--207} (\bibinfo {year}
  {2001})}\BibitemShut {NoStop}%
\bibitem [{\citenamefont {Månsson}\ \emph {et~al.}(1977)\citenamefont
  {Månsson}, \citenamefont {Sellers}, \citenamefont {Stridh},\ and\
  \citenamefont {Sunner}}]{maansson1977}%
  \BibitemOpen
  \bibfield  {author} {\bibinfo {author} {\bibfnamefont {M.}~\bibnamefont
  {Månsson}}, \bibinfo {author} {\bibfnamefont {P.}~\bibnamefont {Sellers}},
  \bibinfo {author} {\bibfnamefont {G.}~\bibnamefont {Stridh}},\ and\ \bibinfo
  {author} {\bibfnamefont {S.}~\bibnamefont {Sunner}},\ }\bibfield  {title}
  {\enquote {\bibinfo {title} {Enthalpies of vaporization of some 1-substituted
  n-alkanes},}\ }\href
  {https://doi.org/https://doi.org/10.1016/0021-9614(77)90202-6} {\bibfield
  {journal} {\bibinfo  {journal} {J. Chem. Thermodyn.}\ }\textbf {\bibinfo
  {volume} {9}},\ \bibinfo {pages} {91--97} (\bibinfo {year}
  {1977})}\BibitemShut {NoStop}%
\bibitem [{\citenamefont {Svensson}(1979)}]{svensson1979}%
  \BibitemOpen
  \bibfield  {author} {\bibinfo {author} {\bibfnamefont {C.}~\bibnamefont
  {Svensson}},\ }\bibfield  {title} {\enquote {\bibinfo {title} {Enthalpies of
  vaporization of 1-decanol and 1-dodecanol and their influence on the
  ch2-increment for the enthalpies of formation},}\ }\href
  {https://doi.org/https://doi.org/10.1016/0021-9614(79)90099-5} {\bibfield
  {journal} {\bibinfo  {journal} {The Journal of Chemical Thermodynamics}\
  }\textbf {\bibinfo {volume} {11}},\ \bibinfo {pages} {593--596} (\bibinfo
  {year} {1979})}\BibitemShut {NoStop}%
\bibitem [{\citenamefont {Kemme}\ and\ \citenamefont
  {Kreps}(1969)}]{kemme1969}%
  \BibitemOpen
  \bibfield  {author} {\bibinfo {author} {\bibfnamefont {H.~R.}\ \bibnamefont
  {Kemme}}\ and\ \bibinfo {author} {\bibfnamefont {S.~I.}\ \bibnamefont
  {Kreps}},\ }\bibfield  {title} {\enquote {\bibinfo {title} {Vapor pressure of
  primary n-alkyl chlorides and alcohols},}\ }\href
  {https://api.semanticscholar.org/CorpusID:95456631} {\bibfield  {journal}
  {\bibinfo  {journal} {J. Chem. Eng. Data}\ }\textbf {\bibinfo {volume}
  {14}},\ \bibinfo {pages} {98--102} (\bibinfo {year} {1969})}\BibitemShut
  {NoStop}%
\bibitem [{\citenamefont {Čenský}\ \emph {et~al.}(2010)\citenamefont
  {Čenský}, \citenamefont {Roháč}, \citenamefont {Růžička},
  \citenamefont {Fulem},\ and\ \citenamefont {Aim}}]{censky2010}%
  \BibitemOpen
  \bibfield  {author} {\bibinfo {author} {\bibfnamefont {M.}~\bibnamefont
  {Čenský}}, \bibinfo {author} {\bibfnamefont {V.}~\bibnamefont {Roháč}},
  \bibinfo {author} {\bibfnamefont {K.}~\bibnamefont {Růžička}}, \bibinfo
  {author} {\bibfnamefont {M.}~\bibnamefont {Fulem}},\ and\ \bibinfo {author}
  {\bibfnamefont {K.}~\bibnamefont {Aim}},\ }\bibfield  {title} {\enquote
  {\bibinfo {title} {Vapor pressure of selected aliphatic alcohols by
  ebulliometry. part 1},}\ }\href
  {https://doi.org/https://doi.org/10.1016/j.fluid.2010.06.019} {\bibfield
  {journal} {\bibinfo  {journal} {Fluid Ph. Equilib.}\ }\textbf {\bibinfo
  {volume} {298}},\ \bibinfo {pages} {192--198} (\bibinfo {year}
  {2010})}\BibitemShut {NoStop}%
\bibitem [{\citenamefont {Rose}, \citenamefont {Papahronis},\ and\
  \citenamefont {Williams}(1958)}]{rose1958}%
  \BibitemOpen
  \bibfield  {author} {\bibinfo {author} {\bibfnamefont {A.}~\bibnamefont
  {Rose}}, \bibinfo {author} {\bibfnamefont {B.}~\bibnamefont {Papahronis}},\
  and\ \bibinfo {author} {\bibfnamefont {E.}~\bibnamefont {Williams}},\
  }\bibfield  {title} {\enquote {\bibinfo {title} {Experimental measurement of
  vapor-liquid equilibria for octanol-decanol and decanol-dodecanol
  binaries.}}\ }\href {https://doi.org/10.1021/i460004a008} {\bibfield
  {journal} {\bibinfo  {journal} {Ind. Eng. Chem. Chem. Eng. Data Ser.}\
  }\textbf {\bibinfo {volume} {3}},\ \bibinfo {pages} {216--219} (\bibinfo
  {year} {1958})},\ \Eprint
  {https://arxiv.org/abs/https://doi.org/10.1021/i460004a008}
  {https://doi.org/10.1021/i460004a008} \BibitemShut {NoStop}%
\bibitem [{\citenamefont {Singh}\ and\ \citenamefont
  {Benson}(1968)}]{singh1968}%
  \BibitemOpen
  \bibfield  {author} {\bibinfo {author} {\bibfnamefont {J.}~\bibnamefont
  {Singh}}\ and\ \bibinfo {author} {\bibfnamefont {G.~C.}\ \bibnamefont
  {Benson}},\ }\bibfield  {title} {\enquote {\bibinfo {title} {Measurement of
  the vapor pressure of methanol-n-decanol and ethanol-n-decanol mixtures},}\
  }\href {https://doi.org/doi/pdf/10.1139/v68-210} {\bibfield  {journal}
  {\bibinfo  {journal} {Can J. Chem.}\ }\textbf {\bibinfo {volume} {46}},\
  \bibinfo {pages} {1249--1254} (\bibinfo {year} {1968})}\BibitemShut {NoStop}%
\bibitem [{\citenamefont {Ambrose}\ and\ \citenamefont
  {Sprake}(1970)}]{ambrose1970}%
  \BibitemOpen
  \bibfield  {author} {\bibinfo {author} {\bibfnamefont {D.}~\bibnamefont
  {Ambrose}}\ and\ \bibinfo {author} {\bibfnamefont {C.~H.~S.}\ \bibnamefont
  {Sprake}},\ }\bibfield  {title} {\enquote {\bibinfo {title} {Thermodynamic
  properties of organic oxygen compounds xxv. vapour pressures and normal
  boiling temperatures of aliphatic alcohols},}\ }\href
  {https://api.semanticscholar.org/CorpusID:95990715} {\bibfield  {journal}
  {\bibinfo  {journal} {J. Chem. Thermodyn.}\ }\textbf {\bibinfo {volume}
  {2}},\ \bibinfo {pages} {631--645} (\bibinfo {year} {1970})}\BibitemShut
  {NoStop}%
\bibitem [{\citenamefont {Ambrose}, \citenamefont {Ellender},\ and\
  \citenamefont {Sprake}(1974)}]{ambrose1974}%
  \BibitemOpen
  \bibfield  {author} {\bibinfo {author} {\bibfnamefont {D.}~\bibnamefont
  {Ambrose}}, \bibinfo {author} {\bibfnamefont {J.~H.}\ \bibnamefont
  {Ellender}},\ and\ \bibinfo {author} {\bibfnamefont {C.~H.~S.}\ \bibnamefont
  {Sprake}},\ }\bibfield  {title} {\enquote {\bibinfo {title} {Thermodynamic
  properties of organic oxygen compounds xxxv. vapour pressures of aliphatic
  alcohols},}\ }\href {https://api.semanticscholar.org/CorpusID:95773528}
  {\bibfield  {journal} {\bibinfo  {journal} {J. Chem. Thermodyn.}\ }\textbf
  {\bibinfo {volume} {6}},\ \bibinfo {pages} {909--914} (\bibinfo {year}
  {1974})}\BibitemShut {NoStop}%
\bibitem [{\citenamefont {Garriga}\ \emph {et~al.}(2004)\citenamefont
  {Garriga}, \citenamefont {Martinez}, \citenamefont {Pérez},\ and\
  \citenamefont {Gracia}}]{garriga2003}%
  \BibitemOpen
  \bibfield  {author} {\bibinfo {author} {\bibfnamefont {R.}~\bibnamefont
  {Garriga}}, \bibinfo {author} {\bibfnamefont {S.}~\bibnamefont {Martinez}},
  \bibinfo {author} {\bibfnamefont {P.}~\bibnamefont {Pérez}},\ and\ \bibinfo
  {author} {\bibfnamefont {M.}~\bibnamefont {Gracia}},\ }\bibfield  {title}
  {\enquote {\bibinfo {title} {Isothermal (vapour+liquid) equilibrium at
  several temperatures of (1-chlorobutane+1-octanol, or 1-decanol)},}\ }\href
  {https://doi.org/https://doi.org/10.1016/j.jct.2003.09.009} {\bibfield
  {journal} {\bibinfo  {journal} {J. Chem. Thermodyn.}\ }\textbf {\bibinfo
  {volume} {36}},\ \bibinfo {pages} {59--69} (\bibinfo {year}
  {2004})}\BibitemShut {NoStop}%
\bibitem [{\citenamefont {Ulbig}, \citenamefont {Klüppel},\ and\ \citenamefont
  {Schulz}(1996)}]{ulbig1996}%
  \BibitemOpen
  \bibfield  {author} {\bibinfo {author} {\bibfnamefont {P.}~\bibnamefont
  {Ulbig}}, \bibinfo {author} {\bibfnamefont {M.}~\bibnamefont {Klüppel}},\
  and\ \bibinfo {author} {\bibfnamefont {S.}~\bibnamefont {Schulz}},\
  }\bibfield  {title} {\enquote {\bibinfo {title} {Extension of the univap
  group contribution method: enthalpies of vaporization of special alcohols in
  the temperature range from 313 to 358 k},}\ }\href
  {https://doi.org/https://doi.org/10.1016/0040-6031(95)02595-2} {\bibfield
  {journal} {\bibinfo  {journal} {Thermochimica Acta}\ }\textbf {\bibinfo
  {volume} {271}},\ \bibinfo {pages} {9--21} (\bibinfo {year}
  {1996})}\BibitemShut {NoStop}%
\bibitem [{\citenamefont {Salimi}\ \emph {et~al.}(2024)\citenamefont {Salimi},
  \citenamefont {Nielsen}, \citenamefont {Pedersen},\ and\ \citenamefont
  {Dantan}}]{salimi2024}%
  \BibitemOpen
  \bibfield  {author} {\bibinfo {author} {\bibfnamefont {M.}~\bibnamefont
  {Salimi}}, \bibinfo {author} {\bibfnamefont {R.~V.}\ \bibnamefont {Nielsen}},
  \bibinfo {author} {\bibfnamefont {H.~B.}\ \bibnamefont {Pedersen}},\ and\
  \bibinfo {author} {\bibfnamefont {A.}~\bibnamefont {Dantan}},\ }\bibfield
  {title} {\enquote {\bibinfo {title} {Squeeze film absolute pressure sensors
  with sub-millipascal sensitivity},}\ }\href
  {https://doi.org/https://doi.org/10.1016/j.sna.2024.115450} {\bibfield
  {journal} {\bibinfo  {journal} {Sensors and Actuators A: Physical}\ }\textbf
  {\bibinfo {volume} {374}},\ \bibinfo {pages} {115450} (\bibinfo {year}
  {2024})}\BibitemShut {NoStop}%
\end{thebibliography}%

\end{document}